# Optoelectronic based Quantum Radar: Entanglement Sustainability Improving at High Temperature

Ahmad Salmanogli[1,2] and Dincer Gokcen*[1]

[1]Faculty of Engineering, Electrical and Electronics Engineering Department, Hacettepe University, Ankara, Turkey

[2]Faculty of Engineering, Electrical and Electronics Engineering Department, Çankaya University, Ankara, Turkey

*Abstract*— **In this study, the main focus is laid on the design of the optoelectronic quantum illumination system to enhance the system performance, such as operation at high temperature and confinement of the thermally excited photons. The optomechanical based quantum illumination system has wieldy been studied, and the results showed that operation at high temperature is so crucial to preserve the entanglement between modes. The main problem is that the mechanical part has to operate with a low frequency with which a large number of thermally excited photons are generated and worsened the entanglement. To solve this problem, we focus on replacing the mechanical part with the optoelectronic components. In this system, the optical cavity is coupled to the microwave cavity through a Varactor diode excited by a photodetector. The photodetector is excited by the optical cavity modes and drives the current flow as a function of incident light drives the Varactor diode at which the voltage drop is a function of current generated by the photodetector. To engineer the system, the effect of some parameters is investigated. One of the critical parameters is the microwave cavity to the photodetector coupling factor ($\mu_c$). Our results indicate that this coupling factor induces a significant difference in the new design as compared to the optomechanical quantum illumination system. At some specific values of the coupling factor, the modes remained completely entangled up to 5.5 K and partially entangled around 50 K.**

## I. INTRODUCTION

In recent years, the idea of using quantum phenomena to enhance the performance of conventional sensors has come to the fore in many applications. Among these applications, radar systems [1]-[4], enhancement of the image resolution [5]-[7], design of quantum illumination systems [8]-[13], quantum communication networks [14]-[15], increase of the plasmonic photodetector responsivity [16], enhancement of the plasmonic system decay rate [17], improvement of Raman signals [18], and so forth can be counted. Primarily, a quantum radar utilizes the entangled microwave photons to enhance the detection, identification, and resolution [1], [4], [9], [10]. In the entangled particles, two particles are linked to each other in such a way that the distance between them is not a concern [5], [19]-[20]. Quantum radars use the fundamental advantage of the classical radars, which is the efficient penetration in clouds and fogs due to microwave photons. Also, quantum radars using entangled microwave photons dramatically enhance the signal to noise ratio, resolution, and probability of detection as compared with the non-entangled photons [1], [2], [9]- [10]. It has been reported that all advantages of the quantum radar are specifically devoted to the entanglement phenomenon. Unfortunately, the entanglement is so fragile and unstable, and also it is generally difficult to create and preserve easily. In other words, the environment noise, specifically thermal noise, can easily destroy the entangled states [12], [17]. The most common problem that the quantum radar system deals with is the propagation of the entangled microwave photons into the atmosphere to detect the target. Designers cannot control the free space temperature, pressure, and interaction between incident light with solid particles. However, one of the crucial factors is the temperature by which the thermally induced photons can easily affect entangled microwave photons associated with low-level energy. Also, the low-level energy of the microwave photon is another factor of the fragility of the entangled states utilized by a quantum radar. This point has been investigated in more detail in some published works [9], [10]. These studies focused on some parameters such as teleportation fidelity [10], signal to noise ratio and error probability [9] in the quantum illumination system. Moreover, the effect of some critical parameters, such as quantum noise and quantum Brownian noise, have been investigated for the entanglement. Another interesting study related to the quantum illumination system has focused on the generation of the entangled microwave photons using Josephson parametric converter (JPC) [11]. In this work, the generated microwave mode is amplified to facilitate the detection process.

The critical point about the system discussed above is the generation of entangled photons at very low temperatures. In [9] and [10], the optomechanical system had been utilized in which the mechanical part frequency is so small. That is the reason, by which the thermally induced photons are dramatically increased and eventually confined the entangled photons.

Therefore, in the studies mentioned, the operational temperature is restricted to 15 mK. In [11], a JPC has been utilized to generate entangled photons due to the related nonlinearity in which the operating frequency is in the range of GHz; nonetheless, the system operating temperature is limited to 7 mK.

In the study, we use the optoelectronics subsystem to generate the entangled photons. The mechanical part from the traditional tripartite system [22]-[26] is removed, and instead, the optoelectronic components are utilized to couple the microwave cavity modes to the optical cavity modes [21]. One of the interesting points related to this system is the high operating frequency of the optoelectronic subsystem. This significantly reduces the thermally induced photons and the associated noises. It is important to note that the entanglement between modes is strongly improved via the utilization of optoelectronic components. Unlike [21], in which the same subsystems had been modeled using optical pressure effect on the photodetector, the system of our interest is analyzed with the canonical conjugate method [27]-[29]. Therefore, in contrast to the former approach [21], the new one provides some degrees of freedom to elaborate on the system completely.

## II.  THEORY AND BACKGROUND

The schematic of the system is illustrated in Fig. 1 in which each stage of the design is depicted. It is clearly shown in Fig. 1a that the laser light is initially coupled to an optical cavity (OC), and consequently, outcoming from OC excites the photodetector (PD). We utilized a quantum dot (QD) photodetector to enhance the current generation [16]. Therefore, the sensitivity of the system can be strongly increased. The current flowing through the PD is a function of the excitation wave frequency. This indicates the current flow highly depends on the OC modes. The current flow induces a voltage drop across the Varactor diode (VD) by which the capacitance of the VD is manipulated. Thus, the VD capacitance is a function of flowing current, which was shaped as a function of OC modes. In other words, the VD capacitance implicitly depends on OC modes. As clearly shown in Fig. 1a, by changing VD capacitance, the MC resonant frequency is manipulated. Therefore, the MC modes will be affected by the OC modes through coupling PD and VD. In this configuration, there are two crucial coupling factors that one can maneuver on them, to manipulate the non-classical correlation between MC and OC modes. One of the factors relates to OC modes coupling to PD by which the current generation rate is defined. Another factor is the coupling between VD and MC mode with which the microwave photons properties are affected. The latter one is a capacitive coupling between the optoelectronic device and MC. As a short conclusion, the entanglement between two modes will be changed by engineering the mentioned coupling factors. The subsystem illustrated in Fig. 1a is the only stage one can manipulate and engineer the entangled microwave photons; in the following stages shown in Figs. 1b, 1c, and 1d, there are no degrees of freedom either to manipulate or to improve the entanglement between modes. Since the MC modes are generated by the coupled systems which are entangled with the OC modes, they are propagated into the atmosphere (attenuation medium) to detect the target. The effects of atmosphere medium are analyzed quantum mechanically through considering the medium as a sequence of the beam splitter (BS) [27], which is schematically depicted in Fig. 1b. Each beam splitter has two inputs as the incident wave ($c_{i-1}$) and thermally generated photons ($b_i$) and two outputs as the noise ($a_{si}$) and the desired output ($c_i$). The thermally induced photons highly depend on the temperature of the atmosphere varying between 190~300 K. It depends on the atmosphere height. The propagation of the entangled photons into the atmosphere drastically affects the entanglement between modes. Consequently, the effect of the scattering from the target is considered (Fig.1c). The propagated signals have some entangled photons as well as separable ones; it is shown that the amount of the entangled photons incident on target are strongly depended on the atmospheric condictions. So the incident fields containing entaged and separable photons interact with the target atom's field. The interaction between two fields, here, is studied via the sequence of BS containing two inputs as thermally induced photons ($b_{ti}$) and $c_{ai}$ as the propagated signals and an output, which is the scattered photons. The scattered photons amplitude and frequency are fully affected by the target's materials properties. In other words, it is the target's field that manipulates the incident photons properties, and especially the entanglement between modes are strictly affected.

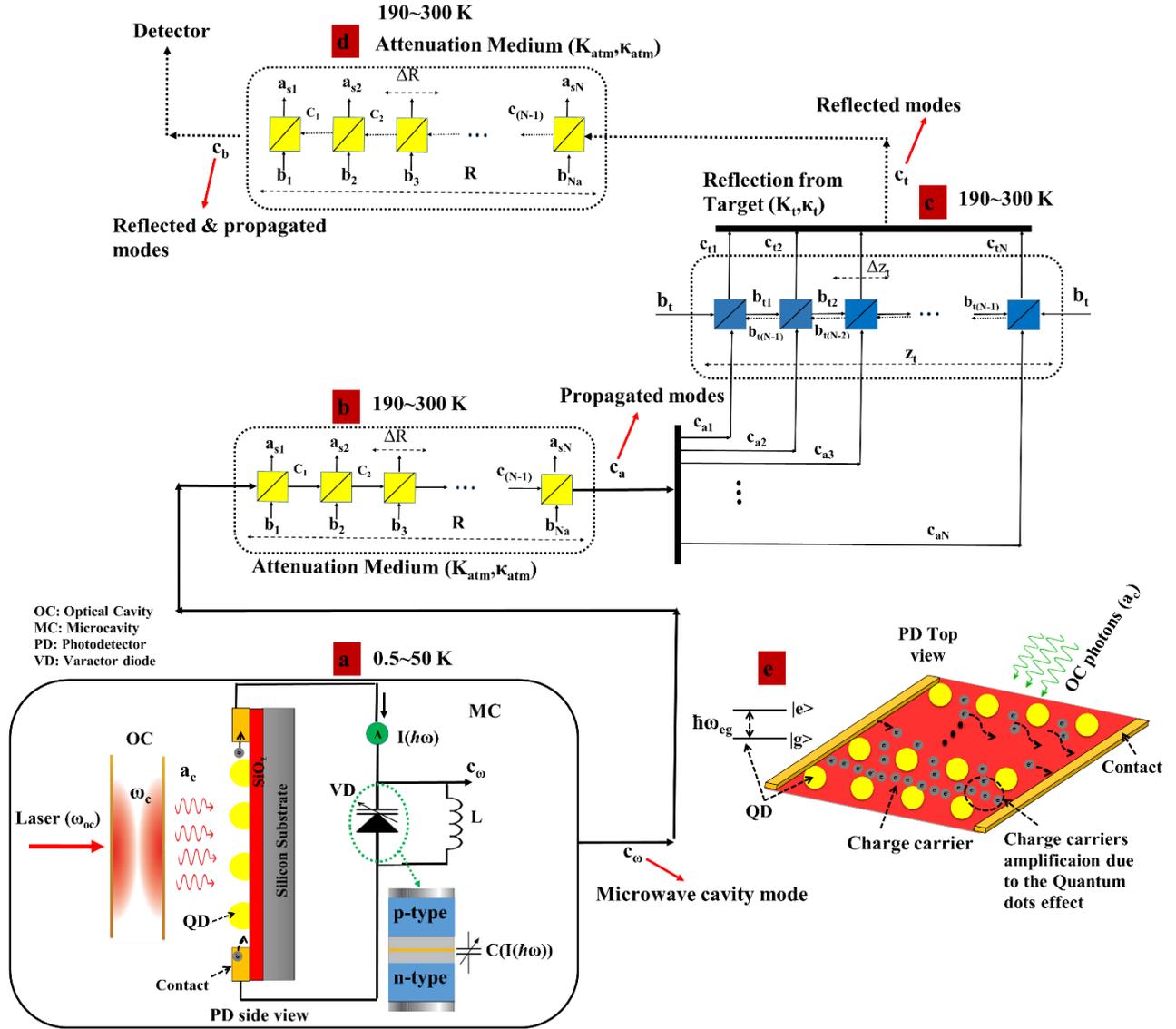

Fig. 1 Schematics of the optoelectronic based quantum illumination system; a) optical cavity coupled to microwave cavity through the optoelectronic device and Varactor diode, b) modeling of the attenuation medium (atmosphere) effect on the microwave cavity modes using BS, c) modeling of the target scattering using BS, d) modeling of the scattering signal transmission through the attenuation medium (atmosphere) using BS and finally the detection, e) photodetector from a top view; it illustrates the interaction of light (photons from OC) with QDs and shows an increment in charge carriers.

Finally, in Fig. 1d, the backscattered photons from the target experience the atmosphere effect in the same way with Fig. 1b. However, there is a critical difference between them; in Fig. 1b the input is $c_\omega$ containing the most entangled photons while in $c_t$, which is the scattered photons, most of the scattered photons lost their entanglement. That means the atmosphere effect on entanglement properties are so crucial in Fig. 1d. Also, the PD top view is depicted in Fig. 1e for better illustration. In this schematic, the QD energy levels and even the effect of QD to generate more carriers are represented.

In the system illustrated in Fig. 1, most of the factors that can affect the entanglement between modes are discussed and modeled using BS. In the following, theoretical analyses of the system will be presented in such a way that one can fully understand the parts or stages of the system in detail.

After a short discussion about the general operation of the system illustrated in Fig. 1, the quantum radar system (subsystem shown in Fig. 1a) is theoretically analyzed using the canonical conjugate method [13]. The total Hamiltonian of the system containing OC, MC, optoelectronics (photodetector-PD), and also the coupling between subsystems given by:

$$H_{OC} = \frac{\varepsilon_0}{2}(E^2 + \omega_c^2 A^2)$$

$$H_{PD} = \frac{P^2}{2m_{eff}} + \frac{1}{2}m_{eff}\omega_{eg}^2 X^2$$

$$H_{MC} = \frac{Q^2}{2C_0} + \frac{\phi^2}{2L} + \frac{C_d v_d}{C_0}Q \quad (1)$$

$$H_{OC-PD} = \alpha_c \frac{AP}{m_{eff}}$$

$$H_{MC-PD} = \frac{-C'(x)}{C_0^2}\left\{\frac{Q^2}{2} + C_d v_d Q\right\}X$$

where $(\mathbf{A}, \mathbf{E})$, $(\mathbf{X}, \mathbf{P})$, and $(\mathbf{\Phi}, \mathbf{Q})$ are the operators for the vector potential and electric field of OC, PD electron-hole position and momentum operators, and MC related phase and charge operators, respectively. Additionally, $\varepsilon_0$, $\alpha_c$, $\omega_{eg}$, $\omega_c$, L, $v_d$, C'(x), $C_0$, $C_d$, and $m_{eff}$ are the free space permittivity, OC-PD coupling coefficient, PD gap-related frequency, OC angular frequency, inductance, MC input driving field, variable capacitor, total capacitor, connection capacitor between $v_d$ and LC circuit, and electron-hole effective mass, respectively.

In the next step, one can define the conjugate variable of the operators $\mathbf{A}$, $\mathbf{X}$, and $\mathbf{\Phi}$ based on the classical definition of the conjugate variables [28]. The contributed Hamiltonian in terms of the creation and annihilation operators is presented as:

$$H_{OC} = \hbar\omega_c \hat{a}_c^+ \hat{a}_c + i\hbar E_c[\hat{a}_c^+ e^{(-j\omega_{oc}t)} - \hat{a}_c e^{(j\omega_{oc}t)}]$$

$$H_{MR} = \frac{\hbar\omega_{eg}}{2}(\hat{P}_x^2 + \hat{q}_x^2)$$

$$H_{MC} = \hbar\omega_\omega \hat{c}_\omega^+ \hat{c}_\omega - j|v_d|C_d\sqrt{\frac{\hbar\omega_\omega}{2C_0}}[\hat{c}_\omega e^{(-j\omega_{o\omega}t)} - \hat{c}_\omega^+ e^{(j\omega_{o\omega}t)}]$$

$$H_{OC-PD} = \hbar\sqrt{\frac{\alpha_c^2\omega_{eg}}{2\varepsilon_0 m_{eff}\omega_c}}(\hat{a}_c^+ + \hat{a}_c)\hat{P}_x$$

$$H_{MC-PD} = -\frac{\hbar\mu_c\omega_\omega}{2d}\sqrt{\frac{\hbar}{\omega_{eg}m_{eff}}}\hat{q}_x \hat{c}_s^+ \hat{c}_s$$

(2)

where $(\mathbf{a_c^+}, \mathbf{a_c})$, and $(\mathbf{c_\omega^+}, \mathbf{c_\omega})$ are the creation and annihilation operators for the optical and microwave cavities, respectively. Also, $\omega_{oc}$ is the input laser angular frequency; $(\mathbf{P_x}, \mathbf{q_x})$ refers to the momentum and position normalized quadrature operator; $E_c$ is the OC cavity input driving rate; d stands for VD capacitor depletion layer width in the steady-state, and $\hbar$ is the reduced Plank's constants. and moreover $\mu_c = C'(x)/C_0$. From MC-PD interaction Hamiltonian, coupling factor is concluded as $g_{o\omega} = (\mu_c\omega_\omega/2d)\times\sqrt{(\hbar/\omega_{eg}m_{eff})}$; this is such an important factor in quantum radar system by which the non-classical correlation between modes can be manipulated. Here in this study, by manipulation of this factor, we intend to preserve the entanglement between modes even at high temperatures. Also, $g_{op} = \sqrt{(\omega_{eg}\alpha_c^2/2\omega_c\varepsilon_0 m_{eff})}$ is the OC-PD coupling rate. In Eq. 2, $H_{oc}$ and $H_{MC}$ contain the field Hamiltonian plus cavity driving by the external source. In this system, using the first-order perturbation theory, one can calculate the coupling coefficient between OC and PD in unit volume as [16]:

$$g_{op} = \frac{\pi\omega_c}{\varepsilon_0 V_m}\mu^2 g_J(\hbar\omega_{eg}).L(\omega_{eg}) \rightarrow$$

$$\alpha_c = \sqrt{\frac{2\omega_c\varepsilon_0 m_{eff}}{\omega_{eg}}} \frac{\pi\omega_c}{\varepsilon_0 V_m}\mu^2 g_J(\hbar\omega_{eg}).L(\omega_{eg}) \quad (3)$$

where $\mu$, $g_J(\hbar\omega_{eg})$, $V_m$, and $L(\omega_{eg})$ are the dipole momentum, PD density of state, volume, and Lorentzian function, respectively. Using Eq. 3 one determines the $\alpha_c$, which depends on the PD properties, so by manipulation of the optical properties of PD, it is possible to change the coupling between OC and PD. We want to use this factor as a critical case for engineering the coupling between cavity modes.

After the calculation of the system total Hamiltonian, the dynamic equation of motion is defined using Heisenberg-Langevin equation. Here, it is necessary to consider the effect of the damping rate and noise due to the interaction of the system with the medium. Also, using rotating wave approximation, the MC and OC are driven at the frequencies $\Delta_\omega = \omega_\omega - \omega_{o\omega}$ and $\Delta_c = \omega_c - \omega_{oc}$, respectively; moreover, PD is driven by $\Delta_{eg} = \omega_{eg} - \omega_c$. Thus, the system dynamic equation of motion is given by:

$$\dot{\hat{a}}_c = -(i\Delta_c + \kappa_c)\hat{a}_c - ig_{op}\hat{P}_x\hat{a}_c + E_c + \sqrt{2\kappa_c}\hat{a}_{in}$$

$$\dot{\hat{c}}_\omega = -(i\Delta_\omega + \kappa_\omega)\hat{c}_\omega + ig_{op}\hat{q}_x\hat{c}_\omega + E_\omega + \sqrt{2\kappa_\omega}\hat{c}_{in} \quad (4)$$

$$\dot{\hat{q}}_x = \Delta_{eg}\hat{P}_x + g_{op}(\hat{a}_c^+ + \hat{a}_c)$$

$$\dot{\hat{P}}_x = -\gamma_p\hat{P}_x - \Delta_{eg}\hat{q}_x + g_{op}\hat{c}_\omega^+\hat{c}_\omega + \hat{b}_{in}$$

where $\kappa_c$, $\gamma_p$, and $\kappa_\omega$ are the optical cavity decay rate, photodetector damping rate, and microwave cavity decay rate, respectively. In this equation, $\Delta_c$, $\Delta_\omega$, and $\Delta_{eg}$ are the related detuning frequencies, and $E_\omega = v_d C_d\sqrt{(\omega_\omega/2\hbar C_0)}$ and $E_c = \sqrt{(2P_c\kappa_c/\hbar\omega_{oc})}$ where $P_c$ is the OC excitation power [22]- [26]. Also, $b_{in}$ is the quantum noise acting on PD, and $a_{in}$ and $c_{in}$ are input noises because cavities interact with the environment.

A simple way to calculate the entanglement for continuous modes is to select a fixed point that OC and MC are driven and worked. Under this condition, the driven field is so strong, and also the coupling constant between OC, PD, and MC, PD are so small; therefore, one can linearize the equation of motion by expanding around the steady-state field (driven field). Thus, it is appropriate to focus on the linearization and calculate the quantum fluctuation around the semi-classical fixed point [22], [26]. To linearize the motion equation, the cavity modes are expressed as the composition of a constant and a fluctuating part as $\mathbf{a_c} = A_s + \boldsymbol{\delta a_c}$, $\mathbf{c_\omega} = C_s + \boldsymbol{\delta c_\omega}$, $\mathbf{q_x} = X_s + \boldsymbol{\delta q_x}$, and $\mathbf{P_x} = P_s + \boldsymbol{\delta p_x}$, where the capital letter with subscript "s" denotes the fixed point, and also $\delta$ indicates the fluctuation part.

In the steady-state condition, for $Re\{A_s\} \gg 1$ and $|C_s| \gg 1$ by which the stability condition is satisfied, the dynamic of motion around the fixed point can be safely linearized. Therefore, the quantum fluctuations of modes around the steady-state points ($A_s$, $P_s$, $X_s$, and $C_s$) are given by:

$$\dot{\delta a_s} = -(i\Delta_c + \kappa_c)\hat{\delta a_c} - ig_{op}\{A_s \hat{\delta p_x} + P_s \hat{\delta a_c}\} + \sqrt{2\kappa_c}\,\hat{\delta a_{in}}$$

$$\dot{\delta c_\omega} = -(i\Delta_\omega + \kappa_\omega)\hat{\delta c_\omega} + ig_{\omega p}\{\hat{\delta q_x} C_s + X_s \hat{\delta c_\omega}\} + \sqrt{2\kappa_{cs}}\,\hat{\delta c_{in}}$$

$$\dot{\delta q_x} = \Delta_{eg}\,\hat{\delta p_x} + g_{op}\{\hat{\delta a_c^+} + \hat{\delta a_c}\}$$

$$\dot{\delta p_x} = -\gamma_p\,\hat{\delta p_x} - \Delta_{eg}\,\hat{\delta q_x} + g_{\omega p}\{C_s \hat{\delta c_\omega^+} + C_s^* \hat{\delta c_\omega}\} + \hat{\delta b_{in}}$$

(5)

The interaction among cavities, e.g., OC with PD and MC with PD can generate CV entanglement, such as the quantum correlation between quadrature operator of the cavity fields [22]-[26]. Therefore, in the following, we need to express the OC and MC fields quadrature fluctuation by a matrix form as:

$$\begin{bmatrix} \dot{\delta q_x} \\ \dot{\delta p_x} \\ \dot{\delta X_c} \\ \dot{\delta Y_c} \\ \dot{\delta X_\omega} \\ \dot{\delta Y_\omega} \end{bmatrix} = \underbrace{\begin{bmatrix} 0 & \Delta_{eg} & \sqrt{2}g_{op} & 0 & 0 & 0 \\ -\Delta_{eg} & -\gamma_p & 0 & 0 & \sqrt{2}g_{\omega p}C_{sr} & -\sqrt{2}g_{\omega p}C_{si} \\ 0 & \sqrt{2}g_{op}A_{si} & -\kappa_c & \Delta_c + g_{op}P_s & 0 & 0 \\ 0 & -\sqrt{2}g_{op}A_{sr} & -\Delta_c - g_{op}P_s & -\kappa_c & 0 & 0 \\ -\sqrt{2}g_{\omega p}C_{si} & 0 & 0 & 0 & -\kappa_\omega & \Delta_\omega - g_{\omega p}q_s \\ \sqrt{2}g_{\omega p}C_{sr} & 0 & 0 & 0 & -\Delta_\omega + g_{\omega p}q_s & -\kappa_\omega \end{bmatrix}}_{A_{i,j}} \times \underbrace{\begin{bmatrix} \delta q_x \\ \delta p_x \\ \delta X_c \\ \delta Y_c \\ \delta X_\omega \\ \delta Y_\omega \end{bmatrix}}_{u(0)} + \underbrace{\begin{bmatrix} 0 \\ \delta b_{in} \\ \sqrt{2\kappa_c}\delta X_c^{in} \\ \sqrt{2\kappa_c}\delta Y_c^{in} \\ \sqrt{2\kappa_\omega}\delta X_\omega^{in} \\ \sqrt{2\kappa_\omega}\delta Y_\omega^{in} \end{bmatrix}}_{n(t)} \quad (6)$$

In Eq. 6, $\delta X_c^{in}$, $\delta Y_c^{in}$, $\delta X_\omega^{in}$, and $\delta Y_\omega^{in}$ are the quadrature operator of the related noises. The solution of Eq. 6 yields to a general form as $u(t) = \exp(A_{i,j}t)u(0) + \int(\exp(A_{i,j}s).n(t-s))ds$, where $n(s)$ is the noise column matrix. The input noises obeying the correlation function [22], [26] and can be expressed as:

$$<a_{in}(s)a_{in}^*(s')> = [N(\omega_c)+1]\delta(s-s')$$
$$<a_{in}^*(s)a_{in}(s')> = [N(\omega_c)]\delta(s-s')$$
$$<c_{in}(s)c_{in}^*(s')> = [N(\omega_\omega)+1]\delta(s-s') \quad (7)$$
$$<c_{in}^*(s)c_{in}(s')> = [N(\omega_\omega)]\delta(s-s')$$
$$<b_{in}(s)b_{in}^*(s')> = [N(\omega_m)+1]\delta(s-s')$$
$$<b_{in}^*(s)b_{in}(s')> = [N(\omega_m)]\delta(s-s')$$

where $N(\omega) = [\exp(\hbar\omega/k_B T_c)-1]^{-1}$ is the equilibrium mean of the thermal photon numbers of different modes, and $k_B$ and $T_c$ are the Boltzmann's constant and system operating temperature (subsystem illustrated in Fig. 1a), respectively. Solving Eq. 6 gives the cavities mode fluctuation, then one can calculate the entanglement between the modes. Here in this system, we specifically focus on the entanglement between OC and MC modes is so critical. Such bipartite entanglement can be quantified through Symplectic eigenvalue given by [30], [31]:

$$\eta = \frac{1}{\sqrt{2}}\sqrt{\sigma \pm \sqrt{\sigma^2 - 2\det(\sigma)}},$$

$$\sigma = \det(A) + \det(B) - 2\det(C) \quad (8)$$

Using Eq. 8, justify that if $2\eta >= 1$, the considered modes are purely separable; otherwise, for $2\eta < 1$, two modes are entangled [31]. Also, in Eq. 8, A, B, and C are 2×2 correlation matrix elements [A, C; $C^T$, B] which can be presented, as an example, for OC-MC modes as:

$$A = \begin{bmatrix} <\delta X_c^2> - <\delta X_c>^2 & 0.5\times <\delta X_c \delta Y_c + \delta Y_c \delta X_c> - <\delta X_c><\delta Y_c> \\ 0.5\times <\delta X_c \delta Y_c + \delta Y_c \delta X_c> - <\delta X_c><\delta Y_c> & <\delta Y_c^2> - <\delta Y_c>^2 \end{bmatrix}$$

$$B = \begin{bmatrix} <\delta X_\omega^2> - <\delta X_\omega>^2 & 0.5\times <\delta X_\omega \delta Y_\omega + \delta Y_\omega \delta X_\omega> - <\delta X_\omega><\delta Y_\omega> \\ 0.5\times <\delta Y_\omega \delta X_\omega + \delta X_\omega \delta Y_\omega> - <\delta Y_\omega><\delta X_\omega> & <\delta Y_\omega^2> - <\delta Y_\omega>^2 \end{bmatrix}$$

$$C = \begin{bmatrix} 0.5\times <\delta X_c \delta X_\omega + \delta X_\omega \delta X_c> - <\delta X_c><\delta X_\omega> & 0.5\times <\delta X_c \delta Y_\omega + \delta Y_\omega \delta X_c> - <\delta X_c><\delta Y_\omega> \\ 0.5\times <\delta Y_c \delta X_\omega + \delta X_\omega \delta Y_c> - <\delta Y_c><\delta X_\omega> & 0.5\times <\delta Y_c \delta Y_\omega + \delta Y_\omega \delta Y_c> - <\delta Y_c><\delta Y_\omega> \end{bmatrix} \quad (9)$$

In order to study the entanglement between OC and MC modes, it is necessary to calculate the expectation value of the quadrature modes operators (e.g., $<\delta X_c>$, $<\delta X_\omega>$, $<\delta X_c^2>$, $<\delta X_\omega^2>$, etc.).

Up to now, the system dynamics equation of motion is analytically derived using the canonical conjugate method and Heisenberg-Langevin equations, and also Symplectic eigenvalue was used as a criterion to analyze the entanglement between two modes. After that, as given in the procedure illustrated in Fig. 1, we have to apply the effect of the attenuation medium, and scattering from a target on the outcoming propagated photons. As the main goal of this study, the subsystem in Fig. 1a should be designed in such a way that other effects such as atmosphere and scattering from the target are not supposed to distort the entanglement between modes completely. In the following, the impact of atmosphere and scattering from the target is briefly investigated from a theoretical perspective. However, one can consider [13] to study in detail about the mentioned effects.

First of all, we considered the effect of the attenuation medium, schematically illustrated in Fig. 1b, as a lossy medium on the entangled modes $c_\omega$ propagating into the medium.

In this study, the atmosphere medium is considered as a package involving the scattering centers. These scattering centers can be modeled with BS using quantum electrodynamic, as shown in Fig. 1b. The modeling starts with the attenuation medium scattering agents in which for $j^{th}$ BS, the input is ($c_{a\omega j}$, $b_j$), and the output is ($c_{a\omega(j+1)}$, $a_{sj}$). Therefore, the continuous form using the attenuation medium's complex coupling coefficient satisfying $|t(\omega)|^2+|r(\omega)|^2 = 1$ where $t(\omega)$ and $r(\omega)$ are the transmission and reflection, can be expressed as:

$$\hat{c}_a(\omega) = e^{[iKatm-\kappa_{atm}(\omega)]R}\hat{c}_\omega(\omega) + i\sqrt{2\kappa_{atm}(\omega)}\int_0^R dz\, e^{[iKatm+\kappa_{atm}(\omega)](R-z)}\hat{b}(\omega,z) \quad (10)$$

where $c_a(\omega)$, $K_{atm}$ and $\kappa_{atm}(\omega)$ are the attenuation medium output mode operator, the real part of the wave vector, and imaginary part of the wave vector, respectively. The scattering agents exponentially attenuate the input mode operator and also introduce the effect of the noise operator expressed in the second part. In the same way, it is possible to examine the entanglement between two modes ($c_a(\omega)$, $a_c$) using Eq. 8.

Another critical factor in a quantum radar is the effect of the reflection from the target, which can destroy the entangled states. In the same way considered for the attenuation medium effect, the quantum electrodynamics is utilized to model the scattering effect. The model to analyze the reflection effect is given in Fig. 1c. The reflection from a target, in the quantum picture, can be defined as the scattering from the target atoms. This process describes the interaction between the electromagnetic quantum field with the quantum field of atoms. During the scattering from the target, the thermally excited photons play an important role through which the entanglement behavior is strongly affected. To describe the process, $j^{th}$ BS is considered as a reflection agent, and the related expression between the input and output operators is given in the continuous form as:

$$\hat{c}_t(\omega) = \left\{ t(\omega)e^{[iKt\Delta z_t]} + 2\kappa_t(\omega)\sqrt{\Delta z_t}\int_0^{z_t} dz\, \{t(\omega)e^{[iKt-\kappa_t(\omega)]}\}^{(z_t-z)} \right\}\hat{c}_a(\omega) + i\sqrt{\kappa_t(\omega)\Delta z_t}\, e^{[iKt-\kappa_t(\omega)]z_t}\hat{b}_t(\omega) \quad (11)$$

where $c_t(\omega)$, $K_t$, and $\kappa_t(\omega)$ are target's scattering output mode operator, the real part of the wave vector, and the imaginary part of the wave vector, respectively. In this equation, the first term handles the direct effect of the target's imaginary part of the dielectric constant, while in the second term, the impact of the thermally excited photons is taken into account. In other words, this equation expresses that the amplitude of the incident wave $c_a(\omega)$ is dramatically decreased. More importantly, the phase raised due to the thermally excited photons in the second term in Eq. 12 strongly distorts the entanglement.

After the calculation of the reflection from the target, the reflected photons are scattered into the atmosphere medium. So, it is necessary to apply the atmosphere effect once again to complete the processes in quantum radar. In this study, we ignore studying the entanglement after detection; however, one can refer to [33],[34] to fully understand about the detection effect and the related quantities such as signal to noise ratio, error probability and so forth.

In the following, the simulation results are detailed, and a comprehensive discussion about how the optoelectronics components in Fig. 1a can enhance the quantum radar system performance to create the entangled photons at high temperature in contrast with a traditional tripartite subsystem utilized in the quantum illumination system [22]-[26].

### III. RESULTS AND DISCUSSIONS

In this section, all of the simulation results are presented and discussed. The data used to model the system illustrated in Fig. 1 is tabulated in Table. 1. First, the effect of the temperature ($T_c$) as a very critical parameter on the entanglement between modes are studied. We considered two-mode entanglement between **$a_c$** & **$c_\omega$** and **$a_c$** & **$c_b$** at $\mu_c = 0.0002$ and $D_{td} = 20$ m where $D_{td}$ is the distance between transmitter and detector. In this study, as shown in Fig. 1, **$a_c$** is the OC mode, **$c_\omega$** is the MC mode, and **$c_b$** is the mode backscattered from the target and before detection. The results are shown in Fig. 2 in which $2\eta$ is depicted versus PD detuning frequency. It is shown that there is an entanglement around $\Delta_{eg}\sim 0$, meaning PD is excited with a frequency $\omega_c$ so close to $\omega_{eg}$. Fig. 2a shows that two modes remain entangled when the temperature $T_c$ reached up to 3500 mK. Actually, there is an objection with the increase of $T_c$ up to 3500 mK at the traditional tripartite system, which has been utilized to generate entangled photons. The problem with increasing $T_c$ up to 3500 mK is that thermally excited photons that can profoundly affect the entanglement between modes.

However, when the MC generated modes are propagated into the atmosphere to detect the target, signal characteristics are not known, and there is actually no control on signals. The effects of the atmosphere and scattering from the target are studied using Eqs. 10 and 11, and the results are shown in Fig. 2b. Accordingly, establishing entangled modes at higher temperature $T_c > 1000$ mK becomes negligible, the backscattering signals turn out to be completely separable. Thus, we should care about the subsystem at which the entangled photons are created. The effect of the atmosphere and scattering from the target can not be controlled; nonetheless, it is possible to define and design a suitable system (Fig. 1a) to generate the entangled states at high temperatures. By this, we intend to subside and decrease the harmful effects of atmosphere and scattering from the target and preserve the entanglement.

Table .1 Constant data parameters used to simulate the system [9]- [10], [22]- [26]

| | |
|---|---|
| $\lambda_c$ | 875 nm |
| $v_d$ | 0.1 mV |
| $\mu_c$ | 0.0002 |
| d | 100 nm |
| $E_{gap}$ | 1.42 ev |
| $\gamma_p$ | 32 ns$^{-1}$ |
| $m_0$ | 9.109383×10$^{-31}$ Kg |
| $m_{eff}$ | 0.45*$m_0$ |
| L | 120 pH |
| $\kappa_c$ | 0.08*$\omega_{ref}$ |
| $\kappa_\omega$ | 0.02*$\omega_{ref}$ |
| $\omega_{ref}$ | 2$\pi$*10$^6$ Hz |
| $C_0$ | 2.67 nF |
| $C_d$ | 2 pF |
| $P_c$ | 30 mW |

One of the critical factors to design such a system is MC-PD coupling factor $g_{\omega p}$ manipulated by $\mu_c$. $g_{\omega p}$ can be managed via other parameters such as d, $m_{eff}$, $\omega_\omega$, and $\omega_{eg}$. The simulations are performed at $T_c = 5000$ mK and $D_{td} = 20$ m to study the entanglement of modes **$a_c$** & **$c_\omega$** (Fig. 3). It is clearly observable that by increasing the coupling factor of MC and PD, the entanglement between states is sharply increased. In fact, this factor creates a significant difference between this work and others [9]-[10], [22], [26] in which the traditional tripartite systems have been used. For better illustration, the critical region of the figure is zoomed in and shown as an inset figure. Eventually, it is possible to design a system to preserve the entangled states when the operating temperature is increased. In the following, it will be shown that the generation of entangled photons at a higher temperature can preserve the entangled states at harsh atmospheric conditions.

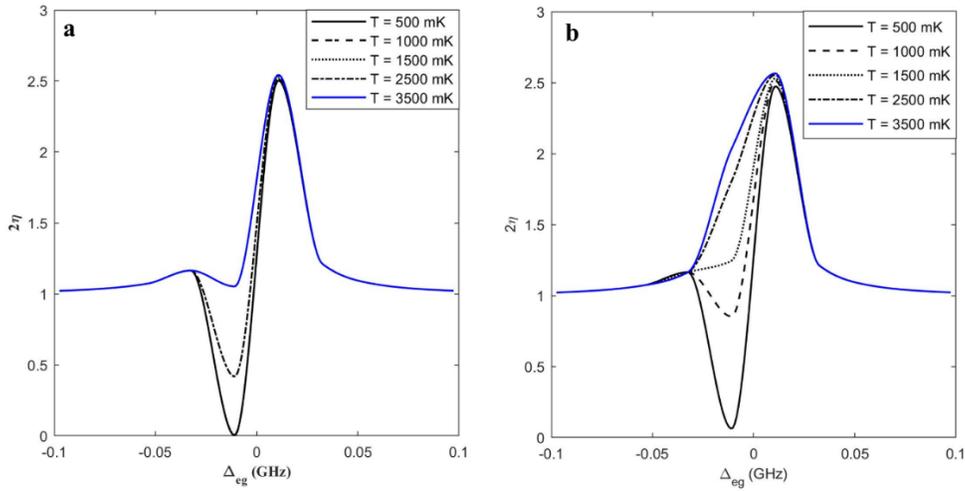

Fig. 2 Temperature effect on entanglement between **(a) $a_c$ & $c_\omega$**, **(b) $a_c$ & $c_b$** at $\mu_c = 0.0002$ and $D_{td} = 20$ m, $\kappa_{atm} = 2*10^{-6}$ 1/m, $\kappa_t = 18.2$ 1/m.

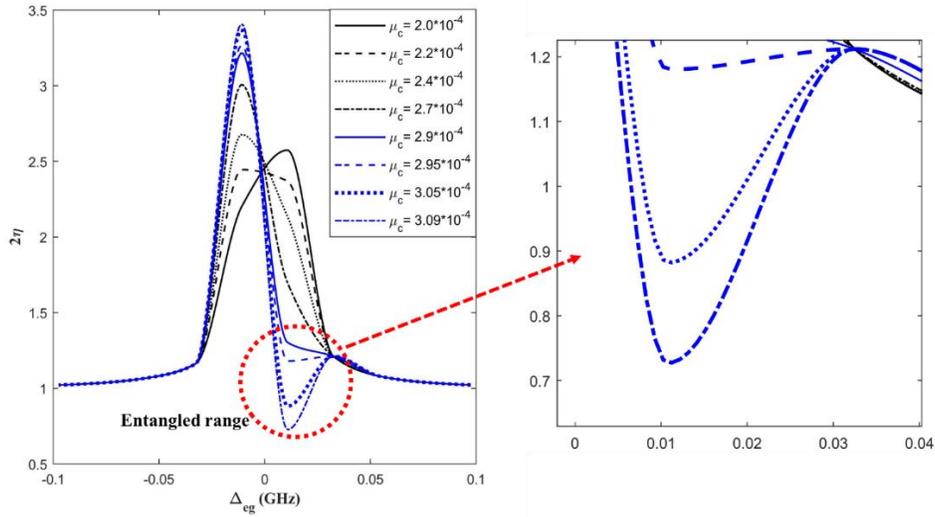

Fig. 3 MC-PD coupling effect on entanglement between $a_c$ & $c_\omega$ at $T_c = 5000$ mK and $D_{td} = 20$ m

The other vital parameter studied in this work is transmitter-detector ($D_{td}$) distance shown in Fig. 4. In this study, the system operating temperature is supposed to be around 1000 mK and for different $D_{td}$, the entanglement between different modes, such as $a_c$ & $c_\omega$, $a_c$ & $c_a$ (attenuation medium effect), $a_c$ & $c_t$ (target's scattering effect), and $a_c$ & $c_b$ (backscattering photons and then propagation into the atmosphere), are studied. According to Eqs. 10 and 11, by increasing the distance between transmitter and receiver, the thermally excited photons will be increased significantly. This leads to decreasing the entanglement between modes. Fig. 4d shows that, when the distance is increased to 2000 m, the entanglements between $a_c$ & $c_a$, $a_c$ & $c_t$, and $a_c$ & $c_b$ leak away entirely, and the input photons for the detector are fully separable. It is contributed to the fragility of the entangled states that are easily affected by the thermally excited photons and noise.

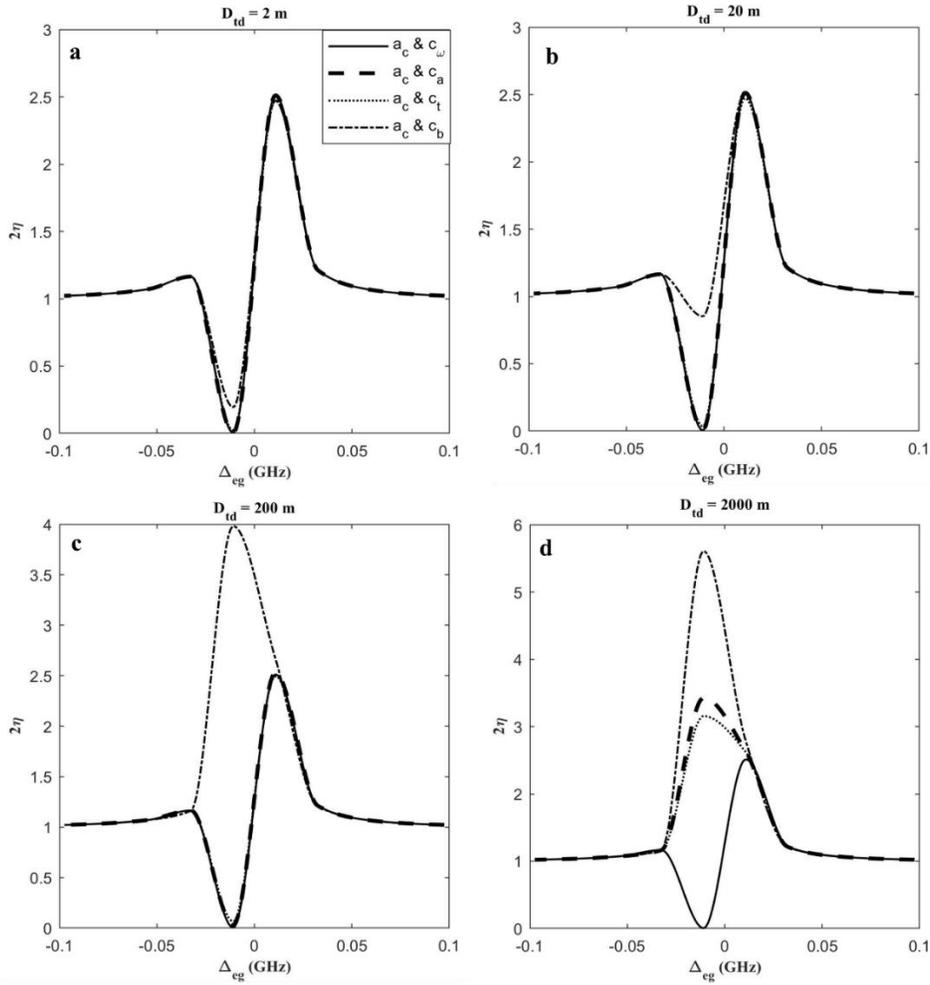

Fig. 4 Transmitter-detector ($D_{td}$) distance effect on entanglement between different modes at $\mu_c = 0.0002$, $T_c = 1000$ mK, $\kappa_{atm} = 2*10^{-6}$ 1/m, $\kappa_t = 18.2$ 1/m.

In the following, we will consider two figures in which the effect of $T_c$ at different modes of quantum radar is studied. These results can be regarded as a significant contribution of this article to the quantum entanglement and quantum illumination systems. The foundation of this study is the design of a system in which entanglement between modes can be preserved at high temperatures. We focused on the MC-PD coupling factor through which one can manipulate the non-classical non-correlation between modes. Fig. 5 reveals that for weak coupling, all modes become separable when $T_c$ is increased up to 3500 mK. It is worthy of mentioning that the MC-PD coupling factor directly affects $a_c$ & $c_\omega$, in which $c_\omega$ is propagated into the atmosphere. Thus, the change of the coupling factor between MC and PD can implicitly manipulate the correlation between modes after propagation into the atmosphere and scattering from the target. To confirm this point, one can consider Fig. 6 in which all simulation conditions are the same as Fig. 5 except $\mu_c$, which controls the coupling strength. According to Fig. 6, especially the backscattering modes $a_c$ & $c_b$, remained entangled even when the temperature is increased up to 50 K and 150 K at some detuning frequencies. Increasing the coupling between MC and PD, the generated modes become strongly non classically correlated to each other, even though the propagation into the atmosphere and scattering from the target cannot force the entanglement to leak away. Therefore, the entanglement between $a_c$ & $c_b$ is just contributed to the entanglement between $a_c$ & $c_\omega$, which is established at a very high temperature.

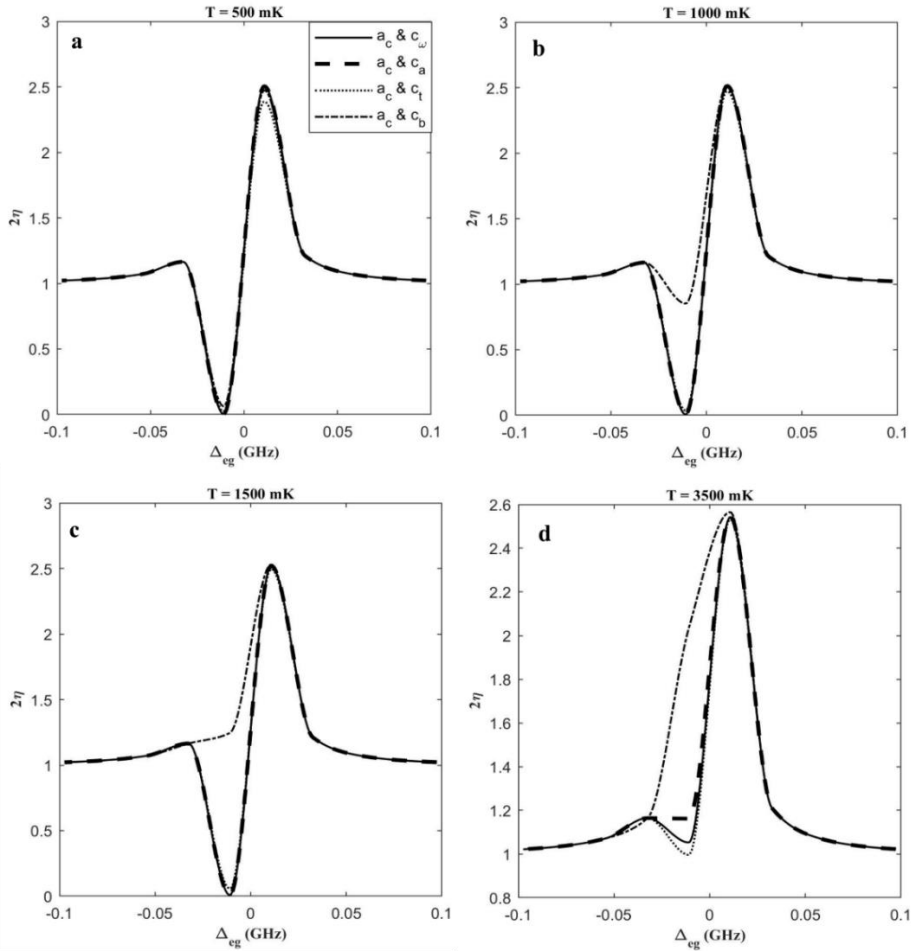

Fig. 5 Temperature effect on entanglement between different modes at $\mu_c = 2*10^{-4}$, and $D_{td} = 20$ m, $\kappa_{atm} = 2*10^{-6}$ 1/m, $\kappa_t = 18.2$ 1/m.

The entangled region in Fig. 6 is depicted with the dashed circle. Moreover, for better illustration, some critical regions of figures are zoomed in and shown as an inset figure. Fig. 6 illustrates that at a relatively high temperature like 50 K, it is possible to generate the entangled photons, and an entangled microwave photon can propagate into the atmosphere, and the entangled photons can also be detected. Optoelectronic components thoroughly contribute to all these exciting features and operating at high temperatures. In this study, the methodology used for coupling between MC and PD is entirely different as compared to the traditional tripartite systems [9-12], which are designed based on the alteration of the distance between the electrodes of the capacitor. In this study, the width of the depletion layer formed in VD determines the capacitance value. The change in the depletion layer width, leading to the shift of $\mu_c$, affects the voltage drop, which is created due to the photocurrent flowing. The factor $\mu_c$ is a critical factor that is manipulated implicitly through the OC modes. While in the traditional tripartite system, the optical pressure generated by OC creates a displacement in capacitor electrodes and induces the change in the capacitance. Additionally, working with a mechanical part confine the operational frequency by which the thermally induced noises can be crucial and limit the operating temperature [9-12].

However, it is evident that increasing $D_{td}$ actively destroys the entanglement between modes. It is because, in the design of the system, we didn't consider an amplifier to intensify the signals before propagation. The amplification of the entangled photons to generate more entangled photons is very critical [11], [32]. It is challenging to achieve both frequency and momentum conservation after the amplification of the photons. So, we preferred to design the quantum radar system without amplifier.

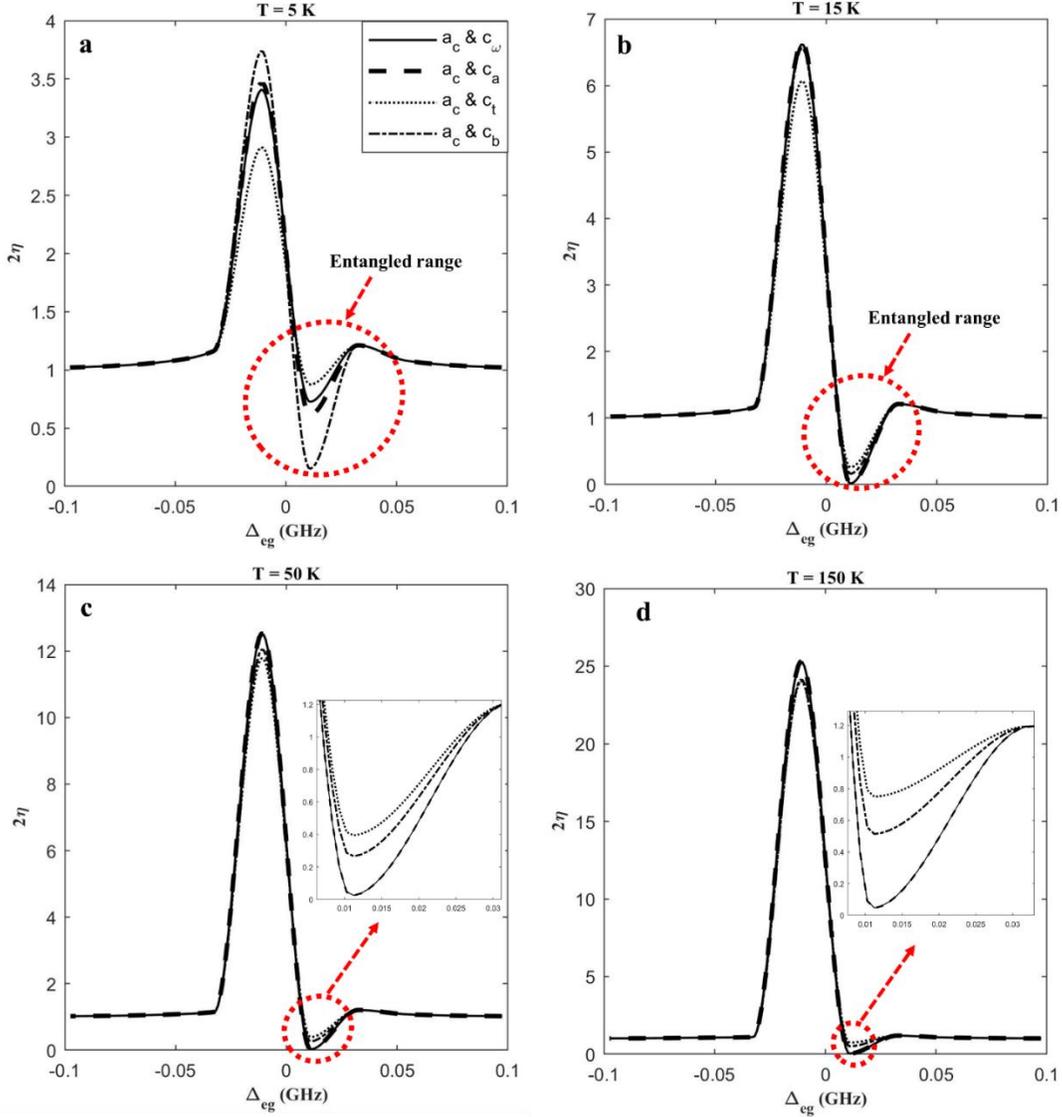

Fig. 6 Temperature effect ($T_c$) on entanglement between different modes at $\mu_c = 3.09*10^{-4}$, and $D_{td} = 20$ m, $\kappa_{atm} = 2*10^{-6}$ 1/m, $\kappa_t = 18.2$ 1/m.

## IV. CONCLUSIONS

In this study, an optoelectronic based quantum radar was designed, modeled, and analyzed. First, an optoelectronic subsystem was designed to generate the entangled photons. This subsystem was the most critical part of our quantum radar, and it was analyzed using the canonical conjugate method. Using this method enabled control over the coupling between MC and PD to manipulate the system to preserve the entanglement between photons. Additionally, this method provided answers to questions such as which parameter specifically subsides the temperature effect on the system and so forth. Following the generation of the entangled microwave photons, it was transmitted into the atmosphere to detect the target. We theoretically derived the effect of the attenuation medium and scattering from the target using a sequential BS system to model the impact of the environment on the incident photons. Finally, entanglements between the different modes were simulated and presented. As a remarkable result, it was shown that one could manipulate the MC-PD coupling factor to effectively subside the temperature effect on the entanglement between OC mode and microwave generated mode. The simulation results revealed that at a temperature around 50 K, it was possible to generate the entangled photons and make them propagate into the atmosphere as an entangled microwave photon as well as the detection of the entangled photons. All these remarkable features, including operation at high temperature, achieved by the optoelectronic subsystem utilized in this study.

**Ahmad Salmanogli** received his B.S. and M.Sc. degrees in Electrical Engineering and Nano&Optoelectronics from Tabriz University. His major research interests are Quantum electronics, Quantum Plasmonic, Plasmonic-Photonic engineering, quantum optics, and plasmonic based nanosensor.

**Dincer Gokcen** (M'08) received the B.S. degree in electrical engineering from Yildiz Technical University in 2005, and Ph.D. degree in electrical engineering from University of Houston in 2010. He has been with National Institute of Standards and Technology, Globalfoundries, and Aselsan, before joining Hacettepe University as an Assistant Professor in 2016. His research interest includes nanofabrication, process engineering technologies, quantum devices, and sensors.